\documentclass[twocolumn,showpacs,preprintnumbers,amsmath,amssymb]{revtex4}

\usepackage{graphicx}
\usepackage{dcolumn}
\usepackage{bm}

\def\zb{zitterbewegung}
\def\Zb{ZITTERBEWEGUNG}
\def\be{\begin{equation}}
\def\ee{\end{equation}}
\def\bea{\begin{eqnarray}}
\def\eea{\end{eqnarray}}

\def\veps{\bf \varepsilon}
\def\ra{\rightarrow}

\begin{document}

\preprint{xxx/yyyy}

\title{Similarity of zitterbewegung of electron to the Adler-Bell-Jackiw anomaly in QED: \\
its observable manifestation in graphene}

\author{S. Arunagiri}

\email{aruna@imsc.res.in}

\affiliation{The Institute of Mathematical Sciences, Chennai 600 113, India}

\date{\today}

\begin{abstract}
The zitterbewegung being proportional to $\sin(\veps t)$, it is depicted as the motion of electron from the positive energy state to that of the negative energy and vice versa in the neighbourhood of Dirac point. Since such transition involves crossing the Dirac point, the same may be the possible realisation and measure of the zitterbewegung. In this respect, we draw a similirity between the zitterbewegung and the Adler-Bell-Jackiw anomaly in (1+1) QED. In a time dependent perturbation theory, the embodiment of zitterbewegung is envisaged in the context of graphene. 
\end{abstract}



\maketitle

\section{Introduction}\label{intro}

Electron in graphene, a monolayer of graphite with honeycomb lattice structure, exhibits relativistic nature as evident from the observed linear dispersion of energy and the zero energy state causing anomalous quantum Hall effect \cite{geim1, geim2, geim3, stormer}. While in the reciprocal lattice space there are two valleys of opposite momentum, $K$ and $K^\prime$, the space of honeycomb lattice consists of two inequivalent sublattices, $A$ and $B$. In the low energy limit, the electron in graphene is described by the Dirac equation: the two valleys are analogue of energy or momentum and the sublattices that of spin. Because of the relativistic-like character of electron in graphene, the phenomenon of zitterbewegung of electron\cite{schrodinger, huang, barut} has been shown renewed interest to unravel its the presence\cite{katsnelson, cserti, rusin, schliemann, winkler, lamata, vaishnav}. 

Schrodinger predicted the zitterbwegung of electron in the Dirac descrption by following the equation of motion for the position operator, ${\bf r}$: 
\be
{d{\bf r} \over {dt}} = c{\bf \alpha} \label{comm1}
\ee
since $[H, {\bf r}] = -i\hbar c {\bf \alpha}$ where  $ H = c {\bf \alpha}\cdot{\bf p} + \beta m c^2$ is free particle Dirac Hamiltonian, ${\bf \alpha}$ and $\beta$ the Dirac matrices, ${\bf p}$ the momentum, $m$ the electron mass and, $c$ the velocity of light. The relation in eq. (\ref{comm1}) is independent of the electron mass. This is in contrast with the corresponding non-relativistic case 
\be
{dx \over {dt}} = {p \over m} \label{nrcomm}
\ee
with the Hamiltonian $H = p^2/2m$. The eq. (\ref{comm1}) is the well-known statement that in special relativity the velocity and momentum are fundamentally different. From the eq. (\ref{comm1}), the position at time $t$ is \cite{schrodinger} 
\be
r = r(0) + c^2 H^{-1} p t + {1 \over 2} i \hbar c \eta_0 H^{-1} e^{-2iHt /\hbar} \label{position}
\ee
where $\eta_0 = \alpha - c H^{-1}p$ which is obtained using the equation of motion for $\alpha$ in the Heisenberg picture and $r(0)$ the position at $t = 0$. In eq. (\ref{position}), the second term is the average value while the third is instantaneous which is oscillating rapidly between the positive and negative energies. This oscillation is known as the zitterbewegung of electron \cite{schrodinger}. The oscillatory behaviour is interpreted to be arising due to interference of positive and negative energy states \cite{sakurai}. Let us note that in Dirac theory, an initial state of mean momentum and spin $(+p, +j)$ given by a Gaussian wavepacket will get evolved in time $t$ into a state that contains an admixture of $(-p, \pm j)$ and $(+p, +j)$ if the wavepacket width is smaller than or equal to the Compton wavelength of the electron. Even if the required confinement is achieved, the oscillatory components is too small to be predictable. Therefore, one looks for a process being an embodiment of the zitterbewegung.  

This article attempts to depict the zitterbewegung phenomenon as the motion of the electron involving transition between the positive and negative energy states near the Dirac point (zero energy level) in energy-momentum space and draw a similarity to the chiral or Adler-Bell-Jackiw (ABJ) anomaly \cite{abj} in (1+1) diemensional QED. Since the transition involves crossing the dirac point, the same may be a possible realisation and measure of zitterbewegung. We formulate such embodiment of zitterbewegung in a time dependent perturbation theory. The formulation is extended to graphene. The claim of similarity is limited to an anomaly-like realisation of the zitterbewegung. It does not imply one can be given in terms of the other.   

\section{Zitterbewegung Vs. ABJ Anomaly}
\label{zbwano}

Let us consider the motion of the electron in the neighbourhood of the Dirac point. An electron state initially of momentum and spin being $(p, s)$ crosses the Dirac point (the zero energy level) from its positive neighbourhood into its negative neighbourhood where it acquires new momentum and spin $(-p, r)$. Continuous occurence of such transition of one state of $(p,s)$ to another of $(-p,r)$ and vice versa in course of time results in the oscillation. Each transition involves crossing of the Dirac point. This happens at a finite interval of time. This oscillation is the zitterbewegung of electron. The periodicity of the oscillation gives the number of the Dirac point crossing. That is the measure of zitterbewegung. The transition does not necessarily change the spin but the momentum. 

The zitterbewegung part of the Dirac current as given in eq. (\ref{current}) (The notation follows the appendix.):
\be
\left< \bar {\psi} \gamma^\mu \psi \right>_Z \propto \sin(2 \varepsilon t + \phi) \label{zbw0}
\ee
where the subscript $Z$ means the zitterbewegung. The rhs in the equation above is the only time dependent part of the current. The period of oscillation depends on the range of values which $\varepsilon, \phi$ and $t$ take. This means the there is a restructuring of $\varepsilon$. This is the feature of zitterbewegung, despite the fact that it is kinematic effect. It is almost impossible for the restructuring to happen of QED vacuum without any perturbation. For, a perturbative analysis is proposed in section \ref{smat}. 

This picture of zitterbewegung naively appears similar to the chiral anomaly in (1+1) QED which is a quantum effect though.

\begin{figure}
\centering%
\includegraphics[scale=0.3]{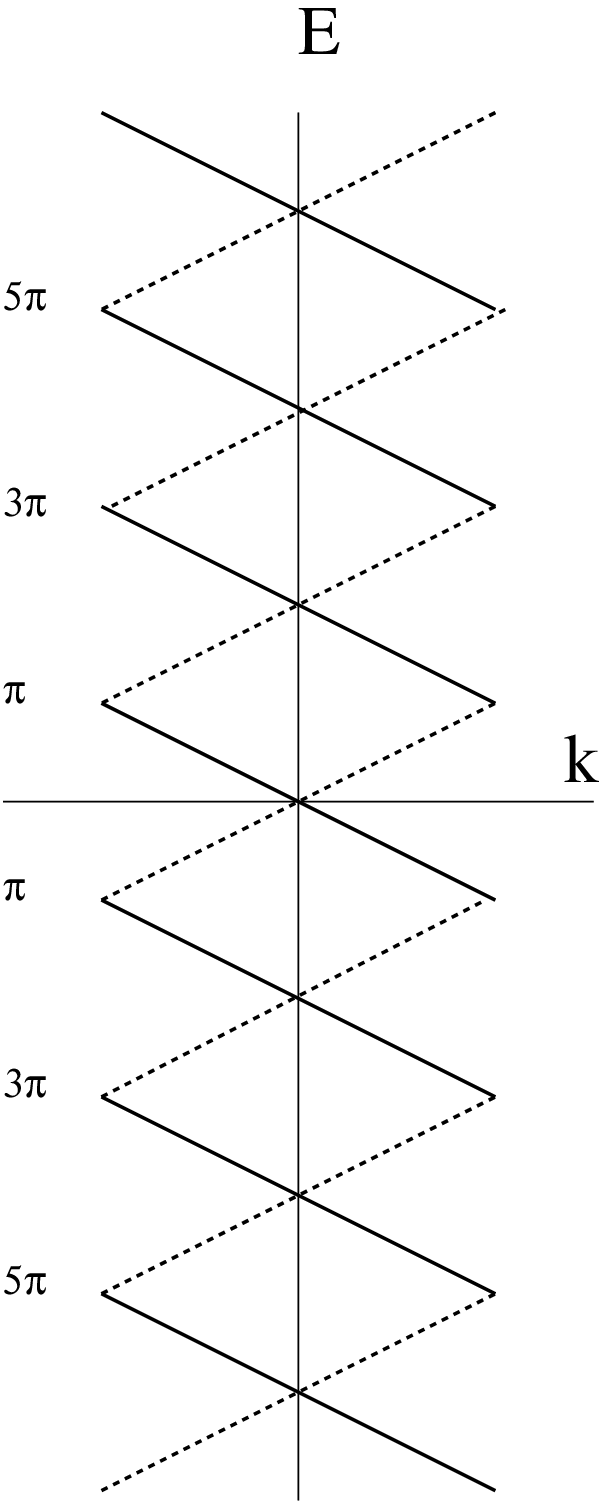}
\caption{Chiral anomaly: shifting of energy levels or transition LH (solid line) and RH (broken line) fermions as a function of $A_1$. Spatial length is set to be unity.}
\label{figure1}
\end{figure}

Let us recall the idea of the ABJ anomaly in (1 + 1) dimensional QED \cite{bertlmann}. The mechanism of the ABJ anomaly relates to the production of Weyl fermions in the presence of an external field, $A_1$. The so-called Weyl fermions are massless spin $\frac{1}{2}$ particles described by a parity non-invariant Hamiltonian. They are of two components, left-handed (RH) and right-handed (RH). In the absence of $A_1$, both the LH and RH fermions meet at the point of degeneracy. On slowly applying the external field, $A_1$, the degeneracy is lifted and the LH fermion moves down below the degenerate level while the RH fermion moves up in the energy-momentum space. For a particular value of $A_1$ both LH and RH fermions meet again at the point of degeneracy. On further application of $A_1$, the degeneracy is lifted again and the LH fermion moves up while the RH fermion moves down. It should be understood as restructuring of energy levels or change of LH fermion into RH and vice versa while crossing the point of degeneracy as shown in Figure \ref{figure1}. The result is the nonconservation of axial charges. That is the ABJ anomaly. The nonconservation of axial charge is seen as the movement electrons from the neighbourhood of LH degeneracy point to that of RH in energy-momentum space, without net production of electrons, in the presence of external field \cite{nielsen}.   

Although the two phenomena are different, the naive similarity arises at the level of restructuring of the energy levels which corresponds to the transition of a state from the positive energy to the negative. It is interesting to note the following: the chiral anomaly (a) occurs due to mass term in the absence of external field (Fujukawa dervation \cite{fujikawa}), (b) chiral current is not conserved and (c) is sensitive to dimensionality, whereas the zitterbewegung (d) involves (although implicitly) spin-orbit interaction, (e) implies non-conservation of spin current and (f) is independent of dimensionality. However, effective dimensional reduction occurs when the system is under external magnetic field. While zitterbewegung is a kinematic effect as a consequence of incorporating special relativity in quantum mechanics, the chiral anomaly is a dynamical phenomenon as a consequence of second quantisation.  From the point of their similarity, we may get an insight of zitterbewegung of electron.

\section{$\bf \Zb$ current}\label{zqft}
\label{smat}
In order to show how an initial state of momentum $p$ and spin $s$ can make a transition into a state of momentum $-p$ and spin $r$ in presence of slowly varying external potential, we consider a time dependent perturbation theory. Embodiment of the zitterbewegung is elucidated in the resulting $S$ matrix.    

For a free massless electron in presence of a slowly varying external potential, $V_\mu(x,t)$.   
\be
c \gamma^\mu p_\mu \Psi(x) = -e\gamma^\mu V_\mu(x,t)\Psi(x)
\ee
The solution to the above equation is, in mode expansion,
\bea
\Psi(x) &=&  \psi(x) \nonumber \\ 
	&+& \sum_{p, s, r} \left[a^+_{p,s}(t) u_p^s e^{ipx} +  a^-_{p,r}(t) v_p^r e^{-ipx} \right] \label{bveq}
\eea
where $\psi(x)$ is the solution of free particle Hamiltonian and can be taken as well defined initial state of $(p, s)$ and 
\bea
a^+_{p,s}(t) &=& -e {1 \over E} \int d^3x^\prime \int^t_{t_i} dt^\prime \bar {u}^s_p \gamma^\mu V_\mu(x^\prime, t^\prime) \psi(x^\prime) e^{-ipx^\prime} \label{aplus}\\
a^-_{p,s}(t) &=& e {1 \over E} \int d^3x^\prime \int^{+t}_{-t} dt^\prime \bar {v}^s_p \gamma^\mu V_\mu(x^\prime, t^\prime) \psi(x^\prime) e^{ipx^\prime} \label{aminus}
\eea
where $E$ is the eigen energy. Of the terms in the second line of eq. (\ref{bveq}), the first term is the probability of the initial positive state remaining the same with $t_i = -\infty$ while the second one is the probability of a positive energy state making transition into a negative energy state with momentum $-p$ and spin $r$ in presence of external field. To make this statement hold, the $t^\prime$ integration in eq. (\ref{aminus}) has to be taken between finite time interval as indicated. Taking one of the limits to be infinity implies creation or annihilation process. The equations in (\ref{aplus}) and (\ref{aminus}) are $S$-matrices of first order in $e$ expansion in presence of external potential.

With the limits of integration specified in equation (\ref{aminus}), the $t^\prime$ integration turns out to be:
\be
\int_{-t}^t e^{-iE t^\prime}dt^\prime = {2 \sin(E t) \over E}
\ee
The scattering matrix corresponding to eq. (\ref{aminus}) is 
\be
S_{u \ra v} = {2e \over {E}} \int d^3x^\prime \bar{v}^r_{-p} \gamma^\mu V_\mu(x^\prime, t^\prime) \psi_n(x^\prime) e^{-i\vec{p} \cdot \vec{x}^\prime} \sin(Et) \label{smg}
\ee
The S-matrix in the above equation can be shown to be containing the current driving the transition of the positive energy state to the negative energy one. For, we take derivative of $S_{u \ra v}$ with respect to the external potential $V_\mu(x, t)$. 

Since we have chosen the potential to be applied adiabatically, we have
\be
V_\mu(x, t) = e^{-\lambda t}V_\mu(x)
\ee
where $\lambda$ is chosen to be a small parameter signifying switch on and off the potential. The variation of $V(x, t)$ with respect to time is negligibly small. Therefore,  $dV = V_0 df(x)$ if $V(x) = V_0 f(x)$ and otherwise $dV = dV_0$. The current is obtained as
\be
j^\mu = E {\partial S_{u \ra v} \over {\partial V_\mu}} \label{curr}
\ee
A simple comparison of the current term given by eq. (\ref{curr}) with the one in eq. (\ref{zbw0}) leads to identification of current given by the scattering matrix, eq. (\ref{curr}), is the zitterbewegung current.  

\section{Graphene Case}
\label{zbgraphene}
We would like to extend the discussion in the previous section to graphene.       

The Dirac equation for a massless electron in graphene in the presence of external magnetic field is 
\be
v_f \gamma^\mu \left(p_\mu-{e \over {c}} A_\mu \right) \psi_n = 0 \label{diracB}
\ee
where $v_f$ is the Fermi velocity of electron in graphene which is $10^{-2} c$, the Greek indices $\mu = 0, 1, 2$, $A_\mu$ = $(A_t, A_x, A_y)$ = $(0, 0, Bx)$ and $p_x$ is expressed in units of $\hbar / l_B$ where $l_B = {\hbar c \over {eB}}$. The energy spectrum is
\be
E_n = \pm \hbar v_f \sqrt{{eB \over {\hbar c}}(2n+1-\sigma_3)}
\ee
The $n = 0$ energy levels are:
\bea
E_0 = \left\{
\begin{array}{l l}
0, & \quad \mbox{$\sigma_3 = +1$}\\
\pm \hbar v_f \sqrt{{2eB \over {\hbar c}}}, & \quad \mbox{$\sigma_3 = -1$}\\ \end{array} \right. 
\eea
The above defines the Dirac point in the presence of applied magnetic field.
The solution of eq. (\ref{diracB}) is 
\be
\psi_n(x) = \exp[i p_y y]\exp\left[-{eB \over {\hbar c}}{(x-x_0)^2 \over {2}} \right] w^j \label{solB}
\ee
where $x_0 = {\hbar c \over {eB}}k_y$ the extension of the wavepacket, and $w^j$, $j = 1, 2, 3, 4$ are normalised spinors.

In the presence of magnetic field the electron states are Landau quantised. The electron is confined into a ``strip" of width $x_0$. Now, We look at the motion of the electron in presence of external electric field, $V_0$ applied adiabatically, parallel to the magnetic field. Then, the $S$-matrix obtained from eq. (\ref{bveq}) is, after the $t^\prime$ integration, 
\be
S = {2e \over E} \int d^3x^\prime \bar{v}^r_{-p} \gamma^0 V_0 \psi_n(x^\prime) e^{-i\vec{p} \cdot \vec{x}^\prime} \sin(E t) \label{number}
\ee
From eq. (\ref{number}) following eq. (\ref{curr}) is the charge density, $j^0(t)$. Obtaining the eigen energy $E = E(B, V_0)$ for $n=0$, we can find the oscillation and dermine the energy restructuring in course of time. As mentioned before in section \ref{zbwano}, the measure of zitterbewegung can be obtained.  

\section{Conclusion}
\label{conclusion}

Although zitterbewegung is a kinematic phenomenon, it is an spin-orbit coupling in an unusual way. The spin-orbit interaction involves both positive and negative energy states. It can be turned into a dynamical when an electron is subjected to an external field. Its signature is the crossing of Dirac point. At this level, its similarity with ABJ anomaly arises as discussed. Despite the similarity, they are not related. However, given the understanding of the chiral anomaly over decades, the anology may be useful to have better understanding of zitterbwegung. In graphene, the spin-orbit interaction is the pseudospin-orbit interaction. Pseudospin is given by the sublattice states. In the desription of graphene with chiral symmetry, the chiral degree of freedom is related to the sublattice states. The chiral symmetry and its breaking in graphene have become interesting aspects of graphene recently \cite{riazuddin, semenoff, hama, kawara}. It is interesting to note the possibility of chiral anomaly induced charge changing transition in graphene \cite{fukushima}. In view of the similarity between the zitterbewegung and the chiral anomaly, we may look for a possible dynamical description of zitterbewegung and realisation in graphene.     
\begin{acknowledgments}

The author is indebted to Prof G. Baskaran, who initiated his interest in zitterbewegung in graphene, for discussions, encouragement and support. He is grateful to Profs R. Simon and R. Shankar for their support and Prof Ramesh Anishetty for useful comments. He thanks the referee for his useful comments and suggestions.   

\end{acknowledgments}

\appendix

\section{Appendix}
\label{append} 

Given the wavefunction for a free massless Dirac particle 
\bea
\psi(x, t) &=& \sum_p \sum_{s = 1, 2} \sqrt{1 \over {2 \varepsilon}} c_{p, s} u^s(p) e^{ipx} + \nonumber\\
	     && \sum_p \sum_{r = 3, 4} \sqrt{1 \over {2 \varepsilon}} d_{p, r}^\dagger v^r(p) e^{-ipx} \label{sol0}
\eea
where $\varepsilon = +\vert pc \vert $ and the normalisation of spinors $\bar {u}^s(p) u^s(p) = \varepsilon$ is used and $px = \vec{p} \cdot \vec{x}/\hbar - \varepsilon t/\hbar$, the current is
\bea
\left< \bar {\psi} \gamma^\mu \psi \right> &=& {1 \over {2 \varepsilon}} c_{p, s}^\dagger c_{p, s} {\bar {u}^s}(p) \gamma^\mu u^s (p) \nonumber\\ 
                                           && + {1 \over {2 \varepsilon}} d_{p, r}^\dagger d_{p, r} {\bar {v}^r}(p) \gamma^\mu v^r (p) \nonumber\\
                                           && + 2 \sum_{p, s, r} {1 \over {2 \varepsilon}} \vert c_{p, s}^\dagger d_{p, r} {\bar {u}^s_p}(p) \gamma^\mu v^r(p) \vert \nonumber\\
                                           && \hskip2.0cm \times \sin(2 \varepsilon t + \phi) \label{current}
\eea
 where $\phi$ is defined to be the phase as
\be
\tan \phi = {Re[c_{p, s}^\dagger d_{p, r} {\bar {u}^s}(p) \gamma^\mu v^r(p)] \over {Im[c_{p, s}^\dagger d_{p, r} {\bar {u}^s}(p) \gamma^\mu v^r(p)]}} \label{phi}
\ee   
In obtaining the form of eq. (\ref{zbw0}), the Hermiticity property of $\gamma^\mu$ is understood.

\end{document}